# Physics from Breit-Frame Regularization of a Lattice Hamiltonian

H. Kröger and N. Scheu

*Département de Physique, Université Laval, Québec, Québec G1K 7P4, Canada.*

*E-mail: hkroger@phy.ulaval.ca, nscheu@phy.ulaval.ca*

(July 2, 1996)

We suggest a Hamiltonian formulation on a momentum lattice using a physically motivated regularization using the Breit-frame which links the maximal parton number to the lattice size. This scheme restricts parton momenta to positive values in each spatial direction. This leads to a drastic reduction of degrees of freedom compared to a regularization in the rest frame (center at zero momentum). We discuss the computation of physical observables like (i) mass spectrum in the critical region, (ii) structure and distribution functions, (iii) $S$-matrix, (iv) finite temperature and finite density thermodynamics in the Breit-frame regularization. For the scalar $\phi_{3+1}^4$ theory we present numerical results for the mass spectrum in the critical region. We observe scaling behavior for the mass of the ground state and for some higher lying states. We compare our results with renormalization group results by Lüscher and Weisz. Using the Breit-frame, we calculate for $QCD$ the relation between the $W^{\mu\nu}$ tensor, structure functions (polarized and unpolarized) and quark distribution functions. We use the improved parton-model with a scale dependence and take into account a non-zero parton mass. In the Bjorken limes we find the standard relations between $F_1$, $F_2$, $g_1$ and the quark distribution functions. We discuss the rôle of helicity. We present numerical results for parton distribution functions in the scalar model. For the $\phi^4$-model we find no bound state with internal parton structure. For the $\phi^3$-model we find a distribution function with parton structure similar to Altarelli-Parisi behavior of $QCD$.

PACS-index: 13.85.-t, 11.10.Ef

## I. INTRODUCTION

The standard model of strong interaction physics ($QCD$) has been confirmed very successfully by comparison between experiment with perturbative and non-perturbative (mostly lattice) calculations. There are a number of observables, which need to be computed non-perturbatively. In some cases non-perturbative computational progress seems very hard to come by. Let us mention the following examples: (a) $S$-matrix for hadron-hadron scattering. (b) Structure functions of the proton, in particular at small values of the Bjorken variable $x_B$. (c) Excited states in the hadron mass spectrum. (d) Finite density thermodynamics of hadronic matter.

In deep-inelastic lepton-hadron scattering ($DIS$) one is interested in structure functions and their interpretation in terms of distribution functions. In order to define distribution functions, one has to specify a reference frame. Possible choices are the rest frame, the infinite momentum frame, light-cone coordinates, or the Breit frame. Conventionally, the infinite momentum frame and light-cone coordinates have been mostly used. The Breit-frame (characterized by $q_0 = 0$) is a distinguished frame in the sense that $Q = \sqrt{-q^2}$, the momentum of the exchanged photon is the resolution ability of the photon to resolve the proton structure. This does not hold in any other frame with $q_0 \neq 0$.

The Breit-frame is not only conceptionally attractive, but we suggest here that is is useful also for non-perturbative numerical calculations. In Ref. [1] we have introduced a new regularization scheme for a lattice Hamiltonian, based on the Breit-frame. Its construction is guided by the kinematical variables which play a rôle in deep-inelastic scattering ($DIS$). We have introduced a momentum lattice based on a the Breit-frame. It is centered around the proton momentum (in the case of $DIS$ proton scattering). The scheme restricts parton momenta to positive values in each spatial direction and links the maximal parton number to the lattice size. This leads to a drastic reduction of degrees of freedom compared to a regularization in the rest frame with the center at zero momentum. In Ref. [1] we have applied the scheme to the scalar $\phi_{3+1}^4$-model. We have computed the mass spectrum and extracted physics close to the critical line (second order phase transition). We found very good agreement with the predictions of the renormalization group by Lüscher and Weisz [2]. To our knowledge, in Ref. [1] critical behavior of a 3 + 1 dimensional field theory has been extracted for the first time from a Hamiltonian formulation.

The successful working of the method with respect to the relative small numerical effort (diagonalization of matrices in the order of 50) makes us cautiously optimistic that other physical observables or other models could be treated as well. In this work we want to elaborate on these ideas. In sect.2 we present the Hamiltonian formulation in the Breit-frame regularization. We explain physical and mathematical reasons for the working of the method. The calculation of the mass spectrum for the $\phi_{3+1}^4$ model with numerical results near the critical line are presented in sect.3. In sect.4 we discuss the structure functions and distribution functions in the Breit-frame. For $QCD$ we compute analytically the relation between the hadronic tensor $W^{\mu\nu}$, the structure functions $F_1$, $F_2$, $g_1$, $g_2$ and the quark distribution functions. We present numerical results for the parton distribution function for the scalar models $\phi_{3+1}^4$ and $\phi_{3+1}^3$. The usefulness of this method eventually depends on its potential in numerical calculations of gauge theories. Thus the Breit-frame regularization for lattice gauge theories is given in sect.5. In sect.6, we discuss advantages of the Breit-frame regularization for the purpose to compute the $S$-matrix of a scattering reaction from a Hamiltonian lattice formulation. Finally, the computation of thermodynamical observables at finite temperature **and** finite density from the Breit-frame is the topic of sect.7. A summary is given in sect.8.

## II. FORMALISM

### A. Hamiltonian formulation

Considering non-perturbative methods in many-body physics, statistical mechanics and field theory, most successful techniques are sum-rule techniques and lattice field theory in the Lagrangian formulation using Monte-Carlo methods to compute functional integrals. Although there is a Hamiltonian formulation of lattice field theory, i.e., the Kogut-Susskind Hamiltonian in lattice gauge theory [3], Hamiltonian methods have not been mainstream in the domain of non-perturbative methods. One of the basic reasons was that in order to treat adequately the physical degrees of freedom, very many virtual particles have to be taken into account. As a function of particle number the dimension of Hilbert space increases exponentially. Nevertheless (maybe due to shortcomings or slow progress in Lagrangian lattice field theory), over recent years several workers have explored Hamiltonian methods, by trying to work with effective Hamiltonians with a small number of degrees of freedom. Examples are the work by Lüscher [4] and van Baal [5], the exp[$S$] method coming from nuclear physics [6], applications of the Kogut-Susskind Hamiltonian to compute glueball masses and string tension in $QCD$ [7], the Hamiltonian approach in light-cone quantization [8], and quite recently a Hamiltonian renormalization group approach [9]. These approaches have employed quite different strategies to cope with the problem of a large number of degrees of freedom: E.g., Wilson and co-workers have pursued the idea of the renormalization group, i.e., thinning out degrees of freedom and constructing a new (renormalized) Hamiltonian with a sufficiently small number of effective degrees of freedom. Lüscher [4] and van Baal [5] have discovered that much physics of the low-lying $QCD$-spectrum, at least for small lattices can be described by zero-momentum dynamics plus a suitable treatment of the remaining degrees of freedom. The idea of the exp[$S$] method [6] is that the linked cluster structure underlying a ground-state in a many-body theory can be generated by a suitable operator $S$, and automatically guarantees the correct infinite volume singularity of the ground state energy. In the applications of the Kogut-Susskind Hamiltonian to $QCD$ [3], several groups have developed clever ways to take into account the high number of plaquettes and closed loop variables, e.g., via the $t$-expansion method by Horn and co-workers [10]. Finally, an advantage of the regularized (discretized) light-cone Hamiltonian method is that light-cone momenta $p^+$ of all partons are positive and add up. Thus a total light-cone momentum $P^+$ drastically limits the number of degrees of freedom. However, this does not hold for the perpendicular momentum $p_\perp$.

Here we pursue the following alternative. Let us consider as example the scalar model given by the Hamiltonian

$$H = \int d^3x \, \frac{1}{2}(\frac{\partial \phi}{\partial t})^2 + \frac{1}{2}(\vec{\nabla}\phi)^2 + \frac{m_0^2}{2}\phi^2 + \frac{g_0}{4!}\phi^4, \tag{1}$$

where $m_0$ and $g_0$ are the bare mass and coupling constant, respectively. We construct the corresponding Hamiltonian in momentum space

$$H = \sum_{\vec{k}} \omega(\vec{k}) a^\dagger(\vec{k}) a(\vec{k}) + \sum_{\vec{k}\vec{l}\vec{m}} \frac{g_o}{4(2\pi)^3} \times \tag{2}$$

$$\left[ 4 \cdot \frac{a^\dagger(\vec{k}) a(\vec{l}) a(\vec{m}) a(\vec{k}+\vec{l}+\vec{m})}{\sqrt{\omega(\vec{k})\omega(\vec{l})\omega(\vec{m})\omega(\vec{k}+\vec{l}+\vec{m})}} + 6 \cdot \frac{a^\dagger(\vec{k}) a^\dagger(\vec{l}) a(\vec{m}) a(\vec{k}+\vec{l}-\vec{m})}{\sqrt{\omega(\vec{k})\omega(\vec{l})\omega(\vec{m})\omega(\vec{k}+\vec{l}-\vec{m})}} + $$

$$4 \cdot \frac{a^\dagger(\vec{k}+\vec{l}+\vec{m})a^\dagger(\vec{l})a^\dagger(\vec{m})a(\vec{k})}{\sqrt{\omega(\vec{k}+\vec{l}+\vec{m})\omega(\vec{l})\omega(\vec{m})\omega(\vec{k})}}+\right]+12g_0\sum_{\vec{k}}\frac{1}{4(2\pi)^3\omega(\vec{k})}a^\dagger(\vec{k})a(\vec{k})\sum_{\vec{l}}\frac{1}{\omega(\vec{l})}.$$

We did not normal order, but we have subtracted the vacuum energy. We have written the Hamiltonian in discretized form by introducing a lattice in momentum space with a momentum resolution $\Delta k$ and a momentum cut-off $\Lambda$. Conventionally one would choose a regular lattice, being symmetrical with respect to momentom zero (rest-frame). This would be suitable to compute the vacuum state with the quantum number $P = 0$. Our alternative is: We choose the same regularization, however, we retain only those lattice momenta which correspond to fast moving partons going in the same direction as the hadron (proton). This is called Breit-frame regularization in the following. As will be shown below this yields a drastic reduction of the effective degrees of freedom compared to the rest frame.

### B. Breit-frame regularization

The most important experiment in order to probe the structure of hadrons is deep inelastic scattering ($DIS$). Its simplest form involves inclusive scattering of an unpolarized lepton off a hadronic target. Let us recall some basic notations [11]. The hadron in its ground state interacts with the probing lepton by the exchange of a virtual photon (or neutrino). The hadron (proton) carries momentum $P$ before the collision and goes over to a hadronic state $X$ with momentum $P_X$. The electron has correspondingly momenta $k$ and $k'$. The exchanged photon carries momentum $q = k - k'$. One defines $Q^2 = -q^2$. In Feynman's parton model it is assumed that the proton is made up of constituents, i.e., the partons. They are weakly bound, i.e., the binding energy is small compared to the resolution ability $Q := \sqrt{-q_\mu q^\mu}$ of the probing photon. Let $p$ denote a parton momentum. Conventionally, one introduces the Bjorken variable $x_B := \frac{Q^2}{2P_\mu q^\mu} = \frac{q \cdot p}{q \cdot P} = \frac{p^{(L)}}{P^{(L)}}$. The superscript $L$ denotes the longitudial direction, i.e. the direction of $\vec{P}$. The second equation results from the impulse approximation, i.e. the partons are on the mass shell. The last identity holds in the Breit-frame. The *Breit frame* of the hadron is defined by the requirements that the photon energy $q_0$ be zero and that the photon momentum $\vec{q}$ be antiparallel to the hadron momentum $\vec{P}$. In this frame, the longitudinal component of the parton momentum obeys $p_L = Q/2$. The rationale for this particular choice of frame is that $QCD$ structure functions $F(x_B, Q)$ can be interpreted as a linear combination of parton momentum distribution functions $f(x_B, Q)$, which have a more intuitive interpretation. This relation holds for *leading twist* (higher twists are suppressed for large $Q^2$). Structure functions are another way of expressing scattering cross sections. The distribution function of a parton counts the number of those partons with a given momentum fraction $x_B$ in the proton. For a precise definition see Ref. [11].

If the hadron is in its ground state, then the longitudinal momentum $p_L$ of the parton can neither be negative nor can it be greater than $P_L$.

$$0 \leq p^{(L)} \leq P^{(L)}. \tag{3}$$

Thus follows the well-known constraint $0 \leq x_B \leq 1$. Eq.(3) can be viewed as a regularization of the longitudinal parton momenta. However, it does not restrict the transverse components of parton momenta. In the Bjorken limit ($Q^2 \to \infty$, $P \cdot q \to \infty$, $x_B = const.$) combined with the Breit-frame the hadron is a fast moving object. In momentum space, an object which is spherical in the rest frame becomes prolate in a fast moving frame. Hence it is physically justified to restrict the transverse parton momenta to a finite region of ellipsoid (prolate) shape. In particular, we have chosen a sphere centered at the mid-point of the interval $[0, P^{(L)}]$ (any prolate ellipsoid lies within this sphere), given by

$$(\vec{p} - \vec{P}/2)^2 \leq (\vec{P}/2)^2. \tag{4}$$

One should note that this constraint also follows directly from $0 \leq x_B \leq 1$ and going into the "parton Breit-frame" (defined by $q^0 = 0$ and $\vec{q}$ being anti-parallel to the *parton* momentum $\vec{p}$) where $x_B$ takes the form $x_B = \frac{\vec{p} \cdot \vec{p}}{\vec{P} \cdot \vec{p}}$.

Because we are working in the Hamiltonian approach we need to define a basis in Hilbert space. We construct the Hilbert space as a Fock space of free particles and select (parton) momenta $\vec{p}$ from a bounded domain corresponding to $DIS$ as given by Eq.(4). This is an *assumption* based on the physical intuition that the experimentally observable parton momenta are those which dominate the quantum dynamics. This assumption has been tested by computing critical behaviour of renormalized masses and a good agreement with analytical scaling behaviour has been observed (see below).

Now we introduce a momentum lattice regularization: In order to have a practically convenient lattice we further constrain the parton momenta from Eq.(4), namely by selecting a regular cube centered at $\vec{P}/2$ and located inside the ball given by Eq.(4). I.e., the parton momenta $\vec{p}$ are restricted to the domain

$$0 \leq p_i \leq \Lambda = \frac{|\vec{P}|}{\sqrt{3}} \quad \text{for } i = x, y, z. \tag{5}$$

Inside this domain we define lattice momenta $\vec{p} := \vec{n}\Delta p$ where $\vec{n}$ is an integer vector and $\Delta p$ is the momentum lattice resolution. One notices that all lattice momenta are non negative. Contrary to regularization in the rest frame which does *not* limit the particle number, our approach has the following important property: The effective Hilbert space is built from the Fock states $|(a^\dagger_{\vec{k}_1})^{n_1} \cdots (a^\dagger_{\vec{k}_N})^{n_N}|0>$, with the conditions that the total momentum be $n_1\vec{k}_1 + \cdots + n_N\vec{k}_N = \vec{P}$, which is located on the surface of the Breit domain, *and* that each parton momentum $\vec{k}_i$ be inside the Breit domain, given by Eq.(5). Thus the regularized Hamiltonian is given by Eq.(2), restricted to the effective Hilbert space.

### C. Reasons for reduction of number of effective degrees of freedom

Why does this regularization scheme lead to a Hamiltonian with a small number of effective degrees of freedom? Firstly, for any given state from the effective Hilbert space, the Fock space particle numbers are bounded, if one considers only non-zero parton momenta. This follows from $n_1\vec{k}_1 + \cdots + n_N\vec{k}_N = \vec{P}$. Thus an upper bound on the total particle number is $\frac{|\vec{P}|}{\Delta p}$. This does, however, not limit the zero-mode particle number. The zero-mode has to be taken into account explicitly. The zero-mode only determines the vacuum expectation value of the field $<\phi>$. In this work we only consider the symmetric phase of the model $<\phi>=0$. In the $\phi^4$ model the vacuum expectation value $<\phi>$ is an order parameter for symmetry breaking and thus the field has fluctuation zero in the infinite volume limit. In this limit it becomes a classical variable. Thus for sufficiently large volume it is justified to set the zero-mode to zero. Besides, in models where the zero-mode can not be dropped, the zero-mode describes only one degree of freedom, which can be treated like a quantum mechanical oscillator. In summary, the ultraviolet cutoff $\Lambda$ given by Eq.(5) implies a total particle number cutoff and thus drastically reduces the dimension of the Hilbert space.

Secondly, if one wishes to compute the mass spectrum of a physical particle, but does not want to compute the vacuum, one has the freedom to choose a reference frame boosted to a momentum $P \neq 0$. As is well known from many-body theory and the exp[S] method, the vacuum state energy has a volume divergence, but the energy of a physical particle state does not have such a divergence. Thus choosing a sector with $P \neq 0$ excludes the vacuum state, but may eventually allow to compute more easily the mass of a physical particle, compared to a computation in the rest frame where the vacuum is present.

### III. MASS SPECTRUM AND CRITICAL BEHAVIOUR OF $\phi^4_{3+1}$ THEORY

Firstly, we need to convince ourselves that the method allows the correct computation of physical observables. We have chosen the scalar $\phi^4_{3+1}$ theory because it is a quite well understood theory which has a second order phase transition, allowing to test our method near a critical point. The Hamiltonian of the $\phi^4$ theory is given by Eq.(2), constrained by the Breit condition (5). It is expressed in terms of free field creation and annihilation operators corresponding to the lattice momenta in the Breit-frame. Because the Hamiltonian and the momentum operators commute, we compute the energy spectrum $E_n$ in a Hilbert space sector of given momentum $\vec{P}$. Since we are not in the rest frame we use the mass-shell condition $M_n := \sqrt{E_n^2 - \vec{P}^2}$ in order to obtain the physical mass spectrum.

It is known [2] that the critical line between the symmetric and the broken phase lies entirely in the region where the bare parton mass squared $m_0^2$ is negative. Hence we cannot build the Fock-space in terms of partons with those masses. As a remedy we have split the bare mass squared $m_0^2 = m_{kin}^2 + m_{int}^2$ into a positive kinetic part $m_{kin}^2$ and an interaction part $m_{int}^2$. The Fock states are built from positive bare masses $m_{kin}$. In numerical calculations close to the critical point shown in Fig.[1] we have chosen, for simplicity, a small positive value. We found that the lower lying physical mass spectrum is not very sensitive to the value of $m_{kin}$ (this is not the case for higher lying masses). A better choice of $m_{kin}$ in our view would be be to take the renormalized mass $m_R$. Although $m_R$ is unknown initially, it can be computed by making an initial guess and then iteratively improving the answer.

We have diagonalized the Hamiltonian on two lattices: $\Lambda/\Delta p = 3$ and $\Lambda/\Delta p = 4$. This would correspond to symmetric lattices $[-\Lambda, +\Lambda]$ of size $7^3$ and $9^3$ nodes, respectively. This results in a very small Hilbert space of only 6 and 21 states, respectively. In order to compare our results to those of Lüscher and Weisz [2] we express the bare parameters $m_0$ and $g_0$ in terms of the parameters $\lambda$ and $\kappa$:

$$m_0^2 = (1 - 2\lambda)/\kappa - 8, \quad g_0 = 6\frac{\lambda}{\kappa^2}. \tag{6}$$

Fig.[1] displays the renormalized mass $m_R$ versus $\kappa$. One observes that our results computed on very small lattices are quite close to the results of Lüscher and Weisz [2]. Masses $M$ computed on the lattice must obey $a < 1/M < L$, where $L$ is the length of the lattice and $a$ denotes the lattice spacing of a space-time lattice. It is related to $\Lambda$ by $\Lambda = \frac{\pi}{a}$. It can be shown from perturbation theory [12,2] that the physical masses close to the critical point obey the following scaling law

$$M \sim C\tau^{1/2}|ln\tau|^{-1/6}, \tag{7}$$

where $\tau := 1 - \kappa/\kappa_{crit}$ and $C$ is a constant (integration constant of renormalization group equations). Since the results of Ref. [2] are based on the solution of the renormalization group equations, this scaling law fits their results. One should note, however, that two different regularizations (this work and that of Ref. [2]) in general correspond to two different critical lines corresponding in general to different values of $\kappa_{crit}$. In Tab. [1] we have displayed our results for the critical points $\kappa_{crit}$ as a function of $\lambda$ and compared our results with those of Ref. [2]. Again, our results are very close to those of Lüscher and Weisz. These results cover a domain of the bare parameter space extending quite far away from the Gaussian fixed point at $\kappa = 1/8, \lambda = 0$.

Another way to test continuum physics is to look at the mass ratios $M_n/M_1$ from the spectrum on the lattice and check if they become independent of the cutoff $\Lambda$ or else independent of the coupling constant $g_0(\Lambda)$ (i.e. they scale). These mass ratios $M_n/M_1$ are shown in Fig.[2]. As can be seen, for a number of states $M_n/M_1 \to const$ in a wide range of $\kappa$-values, i.e. they scale. However, for some states $M_n/M_1$ diverges, i.e., there is no scaling. The physical reason behind this is the following: The $\phi_{3+1}^4$ model describes a gas of partons repelling each other [2]. The spectrum of Fig.[2] shows states dominated by the 1-,2-,3-,4- particle Fock space sectors plus a spectrum of excited (scattering) states. The picture of repulsive two-particle-exchange force is confirmed by observation that the mass of the lowest-lying $n$-body state is larger than $n$-times the mass of the one-body state. The states which scale are just those lowest-lying $n$-body states. The higher-lying part of the spectrum consists of states with more nodes in the wave-function than lattice points, having also a wider range and contributions from higher Fock-state sectors. Because in the calculation corresponding to Fig.[2], the parameters $\Delta p$, $\Lambda$ and the parton number cutoff are all kept fixed, we cannot properly describe these higher-lying states. Consequently, they do not show scaling. When we go to bigger lattices ($\Delta p \to 0$) we then observe (not displayed here) more states which show scaling.

## IV. STRUCTURE FUNCTIONS

### A. Why structure functions in the Breit-frame?

Hadron structure is probed by deep inelastic scattering ($DIS$). Over recent years a great deal of experimental data has been gathered from high energy collider experiments. While perturbative quantum chromodynamics ($QCD$) describes successfully the large $Q^2$ dependence of $DIS$ structure functions, it cannot predict the correct dependence on the Bjorken variable $x_B$. Thus much effort has been devoted to compute quark or gluon distribution functions and proton structure functions from $QCD$ with *non-perturbative* methods. E.g., Martinelli et al. [13] have computed the first two moments of the pion structure function via Monte Carlo lattice simulations. Schierholz and co-workers [14] have recently computed moments of proton and neutron structure functions. These calculations are notoriously difficult. A particular problem is the determination of small $x_B$-behavior from a few moments (for the present status of lattice calculations of structure functions see Ref. [15]). This situation calls for alternative techniques.

Let us briefly outline the reasons for the choice of our method: (i) Structure functions are computed from wave functions. Wave functions are defined in Minkowski space. The Hamiltonian approach offers the advantage of allowing direct computation of Minkowsky space observables. E.g., scattering wave functions for glueball-like states in compact $QED_{2+1}$ have been computed in a Hamiltonian formulation on a momentum lattice [16] (for a review of Hamiltonian lattice methods see [17,7,8]). (ii) The usefulness of a momentum lattice to compute physics close to a critical point has been demonstrated in Ref. [19–24]. (iii) The reason for our choice of the Breit frame has been explained above. However, Hamiltonian methods are known to lead to numerical problems because of the huge number of degrees of freedom involved [25]. To the best of the authors' knowledge nobody has succeeded before to observe scaling behaviour indicating continuum physics in a **(3+1)-dimensional** Hamiltonian lattice formulation.

The Breit frame has a distinct property: Only in this frame the photon momentum transfer $Q$ can be interpreted as resolution ability of the photon. The quark and gluon distribution functions of a proton or a neutron which are measured by $DIS$ show a peak for small $x_B$ even for a moderate resolution $Q$ [26]. This indicates a huge number of partons in the proton, because a system of $n$ identical partons would be peaked at $x_B = 1/n$ for symmetry reasons. The physical reason is that the strong forces which bind the proton can easily create gluons or quark-antiquark pairs. Contrary to a typical non-relativistic problem, particle number is not conserved. Consequently, because of the enormous number of degrees of freedom which are usually associated with a relativistic many-body system, it is almost impossible to calculate quark or gluon distribution functions or mass spectra in a Hamiltonian $QCD$ approach [27]. Our regularization, given by Eq.(5), however, enables us to treat a large number of partons with a *reasonable* numerical effort.

The $Q$ *dependence* of the distribution functions is also a many-particle effect. An intuitive explanation for this dependence is that more partons can be seen inside the proton, if the resolution $Q$ is increased. Partons, however, which are heavy with respect to the forces between them, can be described in a simple constituent model because many-particle effects are negligible. Hence, their distribution functions are neither peaked at $x_B = 0$ nor do they depend considerably on the resolution $Q$. Examples are heavy quarkonia, electromagnetically bound particles (such as atoms) or the $\phi^4_{3+1}$ theory which we are investigating below. The renormalized coupling constant of $\phi^4_{3+1}$ theory is weak everywhere in the critical region and the forces between "partons" are even repulsive [2].

### B. Relation between structure functions and distribution functions in the Breit-frame for $QCD$: unpolarized structure functions

In this section we compute analytically the relation between hadronic tensor, structure functions and distribution functions. Because we work in a fast moving frame and *not* in the infinite momentum frame, we can take into account explicitly a non-zero parton mass. The cross section for deep inelastic lepton-hadron (electron-proton) scattering has the form (following Jaffe's notation [28])

$$d^2\sigma \propto l^{\mu\nu} W_{\mu\nu}, \tag{8}$$

where $l_{\mu\nu}$ denotes the leptonic tensor and $W_{\mu\nu}$ stands for the hadronic tensor. The hadronic tensor can be split into a symmetric part, which corresponds to unpolarized structure functions and an anti-symmetric part, corresponding to the polarized structure functions. The symmetric part can be parametrized in terms of the structure functions $F_1$ and $F_2$,

$$W^{\mu\nu}_{sym} = \left(-g^{\mu\nu} + \frac{q^\mu q^\nu}{q^2}\right) F_1 + \left[\left(P^\mu - \frac{\nu}{q^2}q^\mu\right) \cdot \left(P^\nu - \frac{\nu}{q^2}q^\nu\right)\right] \frac{F_2}{\nu}, \tag{9}$$

where $q^\mu$ is the photon momentum, $P^\mu$ is the proton momentum and $\nu = q \cdot P$. Now we choose as reference frame the Breit-frame: In the Breit-frame the proton momentum is $P^\mu = (E, 0, 0, P_3)$ with $E^2 = P_3^2 + M^2$, $M$ being the proton rest mass. The photon momentum is $q^\mu = (0, 0, 0, -Q)$, $Q$ is defined to be $q^2 = -Q^2$. As a result we find that all components of $W^{\mu\nu}_{sym}$ vanish, except for

$$W^{00}_{sym} = -F_1 + \frac{E^2}{P_3 Q} F_2,$$
$$W^{11}_{sym} = W^{22}_{sym} = F_1. \qquad (10)$$

The hadronic tensor is defined [28] as

$$4\pi W^{\mu\nu} = \sum_X (2\pi)^4 \delta(P + q - P_X) <PS \mid J^\mu(0) \mid X><X \mid J^\nu(0) \mid PS>, \qquad (11)$$

where $X$ denotes the unobserved fragments of the proton, $P$ is the proton momentum. We have normalized the proton state to $<P' \mid P> = 2E(2\pi)^3 \delta^3(P' - P)$. $S$ is the proton spin (Pauli-Lubanski vector) normalized to $S^2 = -M^2$. $J^\mu(x) = \bar{\psi}(x)\gamma^\mu \psi(x)$ denotes the fermionic (quark) current. The hadronic tensor can be expressed as a current commutator,

$$4\pi W^{\mu\nu} = \int d^4 y \exp[-iq \cdot y] <P \mid [J^\mu(y), J^\nu(0)] \mid P>. \qquad (12)$$

In deep inelastic scattering it is customary to use the impulse approximation. The partons lie on the mass shell. Thus one can expand the field $\psi(x)$

$$\psi(x) = \sum_s \int \frac{d^3 k}{(2\pi)^{3/2}} (2\omega(\vec{k}))^{-1/2} \left[ u_s(\vec{k}) e^{-ik \cdot x} b_s(\vec{k}) + v_s(\vec{k}) e^{ik \cdot x} d^\dagger_s(\vec{k}) \right]. \qquad (13)$$

The spinors are normalized to $\bar{u}_s u_{s'} = 2m\delta_{s,s'}$, $\bar{v}_s v_{s'} = -2m\delta_{s,s'}$, where $m$ is the parton rest mass. The parton spin is normalized to $s^2 = -m^2$. The creation and annihilation operators obey $[b_s(\vec{k}), b^\dagger_{s'}(\vec{k}')]_+ = \delta_{s,s'} \delta(\vec{k} - \vec{k}')$ and $[d_s(\vec{k}), d^\dagger_{s'}(\vec{k}')]_+ = \delta_{s,s'} \delta(\vec{k} - \vec{k}')$.

In the computation of the matrix element of the current commutator the following leptonic tensor occurs:

$$\begin{aligned} l^{\mu\nu}_{\bar{u}u\bar{u}u}(k, s, k', s') &= \bar{u}(k,s)\gamma^\mu u(k',s')\bar{u}(k',s')\gamma^\nu u(k,s) \\ &= k^\mu k'^\nu + k'^\mu k^\nu + g^{\mu\nu}(m^2 - k \cdot k') - im\epsilon^{\mu\nu\alpha\beta}(k - k')_\alpha (s + s')_\beta, \end{aligned} \qquad (14)$$

and summing over $s'$ yields

$$l^{\mu\nu}_{\bar{u}u\bar{u}u}(k, k', s) = 2\left[ k^\mu k'^\nu + k'^\mu k^\nu + g^{\mu\nu}(m^2 - k \cdot k') - im\epsilon^{\mu\nu\alpha\beta}(k - k')_\alpha s_\beta \right], \qquad (15)$$

being the standard result [28]. Due to the current commutator, there are four fermion fields involved, which gives 16 combinations of fermion and anti-fermion creation and annihilation oprators. A straightforward but lengthy calculation gives the following term (other terms see below)

$$W^{\mu\nu}_{b\dagger bb\dagger b} = \frac{E}{4} \int d^3 k \frac{\delta(q^0 + k^0 - k'^0)}{\omega(\vec{k})\omega(\vec{k} + \vec{q})} \sum_s l^{\mu\nu}_{\bar{u}u\bar{u}u}(k, k+q, s) \Pi^b_s(P, \vec{k}). \qquad (16)$$

Here we have switched to the following normalization of the proton state $<P' \mid P> = \delta^3(P' - P)$. In the calculation occurs the the matrix element $<PS \mid b^\dagger_s(\vec{k}) b_{s'}(\vec{k}') \mid PS>$ which allows to split off $\delta(\vec{k} - \vec{k}')$ due to total three-momentum conservation as well as to split off $\delta_{s,s'}$ due to conservation of spin quantum numbers in the helicity basis. Thus we have defined $\Pi^b_s(PS, \vec{k})$ by

$$<PS \mid b^\dagger_{s'}(\vec{k}') b_s(\vec{k}) PS> = \Pi^b_s(PS, \vec{k}) \delta(\vec{k} - \vec{k}') \delta_{s,s'}. \qquad (17)$$

which is the the expectation value in the proton state of the quark number operator corresponding to momentum $\vec{k}$ and spin $s$. Note that $\Pi^b$ has the same dimension as $<PS \mid PS>$. Now we go into the Breit-frame. In particular, we employ the Breit-condition, Eq.(4) and (5). In the Breit-frame one has $q^0 = 0$, moreover $\vec{k}$ lies on the mass shell, $(k^0)^2 = \vec{k}^2 + m^2$, due to the impulse approximation. Also $\vec{k}' = \vec{k} + \vec{q}$ is on-shell. Thus we obtain

$$W^{\mu\nu}_{b\dagger bb\dagger b} = \frac{E}{4} \int d^2 k_\perp \sum_s \frac{l^{\mu\nu}_{\bar{u}u\bar{u}u}(k_Q, k'_Q, s)}{Q\omega(\vec{k}_Q)} \Pi^b_s(P, \vec{k}_Q), \qquad (18)$$

where we have defined $\vec{k}_Q = (\vec{k}_\perp, Q/2,)$ and $\vec{k}'_Q = (\vec{k}_\perp, -Q/2)$, (parallel and perpendicular denotes the orientation of components with respect to the space-component of the proton momentum).

A second term, which contributes to the $W^{\mu\nu}$ tensor is the following

$$W^{\mu\nu}_{dd^\dagger dd^\dagger} = \frac{E}{4}\int d^2k_\perp \sum_s \frac{l^{\mu\nu}_{\bar{v}v\bar{v}v}(k'_Q, k_Q, s)}{Q\omega(\vec{k}'_Q)} \Pi^d_s(P, \vec{k}_Q). \tag{19}$$

$\Pi^d_s(PS, \vec{k})$ is defined in analogy to Eq.(17), but for the anti-quark number operator. The leptonic tensor $l^{\mu\nu}_{\bar{v}v\bar{v}v}$ corresponding to the $v$-spinor is defined by

$$l^{\mu\nu}_{\bar{v}v\bar{v}v}(k, s, k', s') = \bar{v}(k, s)\gamma^\mu v(k', s')\bar{v}(k', s')\gamma^\nu v(k, s), \tag{20}$$

yielding, after summation over the spin $s'$

$$l^{\mu\nu}_{\bar{v}v\bar{v}v}(k, k', s) = 2\left[k^\mu k'^\nu + k'^\mu k^\nu + g^{\mu\nu}(m^2 - k \cdot k') + im\epsilon^{\mu\nu\alpha\beta}(k - k')_\alpha s_\beta\right]. \tag{21}$$

All other terms give vanishing contributions to $W^{\mu\nu}$ due to the fact that all parton momenta lie in the Breit-sphere.

Thus the symmetric part of $W^{\mu\nu}$, corresponding to the unpolarized structure functions yields the following result in the Breit-frame. The only non-zero components are those with $\mu = \nu = 0, 1, 2$.

$$W^{\mu\mu}_{sym} = \frac{E}{4}\int d^2k_\perp \frac{l^{\mu\mu}_{unpol}(k_Q, k'_Q)}{Q\omega(\vec{k}_Q)} \sum_s \left[\Pi^b_s(P, \vec{k}_Q) + \Pi^d_s(P, \vec{k}_Q)\right], \tag{22}$$

where $l^{\mu\nu}_{unpol}$ denotes the symmetric part of $l^{\mu\nu}_{\bar{v}v\bar{v}v}$ and $l^{\mu\nu}_{\bar{v}v\bar{v}v}$

$$l^{\mu\nu}_{unpol}(k, k') = 2\left[k^\mu k'^\nu + k'^\mu k^\nu + g^{\mu\nu}(m^2 - k \cdot k')\right]. \tag{23}$$

In particular one has

$$\begin{aligned} l^{00}_{unpol}(k_Q, k'_Q) &= 4[m^2 + (\vec{k}_\perp)^2], \\ l^{11}_{unpol}(k_Q, k'_Q) &= 4(k_1)^2 + Q^2, \\ l^{22}_{unpol}(k_Q, k'_Q) &= 4(k_2)^2 + Q^2. \end{aligned} \tag{24}$$

Let us now consider the Bjorken limes of those expressions. The Bjorken limes is defined by $Q \to \infty$ and $x_B = const$. In the Breit-frame, this implies for the proton momentum $P_3$ that $P_3 \to \infty$ and $Q/P_3 = 2x_B = const$. Thus we compute the kinematical factors $E = \sqrt{M^2 + P_3^2} \sim Q/2x_B$ and

$$\frac{l^{00}_{unpol}(k_Q, k'_Q)}{Q\omega(\vec{k}_Q)} = \frac{4[m^2 + (\vec{k}_\perp)^2]}{Q\sqrt{Q^2/4 + (\vec{k}_\perp)^2 + m^2}} \sim Q^{-2} \to 0. \tag{25}$$

This implies

$$W^{00}_{sym} \to 0. \tag{26}$$

On the other hand, Eq.(10) yields in the Bjorken limes

$$W^{00} = -F_1 + \frac{E_P^2}{P_3 Q} F_2 \to -F_1 + \frac{1}{2x_B} F_2. \tag{27}$$

Thus the last two equations imply in the Bjorken limes that the Callan-Gross relation holds

$$2x_B F_1 = F_2, \tag{28}$$

which is the standard result as in the parton model. Similarly we compute

$$\frac{l^{11}_{unpol}(k_Q,k'_Q)}{Q\omega(\vec{k}_Q)} = \frac{4(k_1)^2 + Q^2}{Q\sqrt{Q^2/4 + (\vec{k}_\perp)^2 + m^2}} \to 2,$$
$$\frac{l^{22}_{unpol}(k_Q,k'_Q)}{Q\omega(\vec{k}_Q)} = \to 2. \tag{29}$$

Thus we obtain the following result for $W^{00}_{sym}$, $W^{11}_{sym}$ and $W^{22}_{sym}$, in the Bjorken limes

$$W^{00}_{sym} = 0,$$
$$W^{11}_{sym} = W^{22}_{sym} \to \frac{P_3}{2} \int d^2 k_\perp \sum_s \left[ \Pi^b_s(P,\vec{k}_Q) + \Pi^d_s(P,\vec{k}_Q) \right]. \tag{30}$$

Our regularization scheme allows to compute directly the parton distribution function. We define

$$f(p_3, P_3, \mu) = \int d^2 k_\perp \sum_s \left[ \Pi^b_s(P,\vec{k}_Q) + \Pi^d_s(P,\vec{k}_Q) \right], \tag{31}$$

where $p_3$ is the longitudinal parton momentum, $P_3$ is the longitudinal proton momentum and $\mu$ is a fixed but arbitrary scale parameter with dimension of mass (e.g., $\Lambda_{QCD}$). $f(p_3, P_3, \mu)$ is the probabilty of finding a parton with longitudinal momentum $p_3$ in a boundstate (proton) with longitudinal momentum $P_3$, where momenta are measured in terms of the scale $\mu$. The $W^{\mu\nu}$ tensor is dimensionless. Thus we have

$$W^{22}_{sym} = \frac{P_3}{2} f. \tag{32}$$

Thus $G(p_3, P_3, \mu) = P_3 f(p_3, P_3, \mu)$ is a dimensionless function which scales

$$G(\lambda p_3, \lambda P_3, \lambda \mu) = G(p_3, P_3, \mu). \tag{33}$$

We have shown in the Bjorken limes that $W^{22}_{sym} = F_1$, thus

$$F_1(x, Q, \mu) = \frac{P_3}{2} f(p_3, P_3, \mu). \tag{34}$$

Expressing $p_3$ and $P_3$ in terms of $Q$ and $x$ yields $p_3 = Q/2$ and $P_3 = Q/2x$ and hence

$$F_1(x, Q, \mu) = \frac{Q}{4x} f(Q/2, Q/2x, \mu). \tag{35}$$

Making a scale transformation by multiplying all variables of dimension mass by $\lambda$, where $\lambda$ is chosen to obey $\lambda P_3 = 1$, yields

$$F_1(x, Q, \mu) = \frac{1}{2} f(x, 1, 2x\mu/Q),$$
$$F_2(x, Q, \mu) = x f(x, 1, 2x\mu/Q). \tag{36}$$

Here $f(x, 1, 2x\mu/Q)$ denotes the probabilty of finding a parton with longitudinal momentum fraction $x$ from total longitudinal momentum $= 1$, where the scale is given by $2x\mu/Q$. Note that in Eq.(30) the sum runs over all spin values. For a spin 1/2 parton this is equivalent to a sum over the helicity quantum numbers $+$ and $-$. If we take into account $e_i$, the electric charge of quark of flavor $i$ relative to the charge of the electron, and redefine Eq.(31) by $f = \int d^2 k_\perp \sum_s \Pi^b_s$ and $\bar{f} = \int d^2 k_\perp \sum_s \Pi^d_s$ we obtain from Eq.(36)

$$F_1(x,Q) = \sum_i e_i^2 \frac{1}{2} \left[ f^{(i)}_+(x,1,2x\mu/Q) + f^{(i)}_-(x,1,2x\mu/Q) + \bar{f}^{(i)}_+(x,1,2x\mu/Q) + \bar{f}^{(i)}_-(x,1,2x\mu/Q) \right]$$
$$F_2(x,Q) = \sum_i e_i^2 x \left[ f^{(i)}_+(x,1,2x\mu/Q) + f^{(i)}_-(x,1,2x\mu/Q) + \bar{f}^{(i)}_+(x,1,2x\mu/Q) + \bar{f}^{(i)}_-(x,1,2x\mu/Q) \right]. \tag{37}$$

The standard expression from the parton model [11,28] is given by

$$F_1(x) = \sum_i e_i^2 \frac{1}{2} \left[ q_+^{(i)}(x) + q_-^{(i)}(x) + \bar{q}_+^{(i)}(x) + \bar{q}_-^{(i)}(x) \right]$$
$$F_2(x) = \sum_i e_i^2 x \left[ q_+^{(i)}(x) + q_-^{(i)}(x) + \bar{q}_+^{(i)}(x) + \bar{q}_-^{(i)}(x) \right]. \tag{38}$$

In the Bjorken limes our result, Eqs.(37), agrees with the standard result, Eq.(38). As can be seen, the quark distribution function $q(x)$ occuring in Eq.(38) does not have any $Q$ dependence. It corresponds to the naive parton model, which has no $Q$ dependence. However, perturbative $QCD$ introduces a $Q$ dependence via logarithmic corrections (violation of scaling). Thus one arrives at $q(x, Q, \mu)$ which is a quark distribution function from a "renormalization group improved parton model". $q(x, Q, \mu)$ is interpreted as the probability to find a parton with momentum fraction $x$ in a hadron with momentum $= \infty$, where the resolution (by the photon) $Q$ is finite, and momenta are measured in terms of a mass scale $\mu$. Note that our distribution function $f(x, 1, 2x\mu/Q)$ has a different interpretation than $q(x, Q, \mu)$: $f(x, 1, 2x\mu/Q)$ corresponds to the Breit frame where the hadron moves fast but with *finite* momentum, while $q(x, Q, \mu)$ corresponds to the *infinite* momentum frame. However, in the Bjorken limes both coincide.

### C. Relation between structure functions and distribution functions in the Breit-frame for $QCD$: polarized structure functions

The anti-symmetric part of the $W^{\mu\nu}$ tensor describing the spin dependent part can be parametrized in terms of the spin structure functions $g_1$ and $g_2$ [28]

$$W_{as}^{\mu\nu} = -i\epsilon^{\mu\nu\sigma\rho} q_\sigma \left[ \frac{S_\rho}{\nu} (g_1 + g_2) - \frac{q \cdot S P_\rho}{\nu^2} g_2 \right]. \tag{39}$$

Here $S$ denotes the proton spin, with $P \cdot S = 0$ and $S^2 = -M^2$. The proton spin can be polarized in two ways: $\vec{S} \parallel \vec{P}$ (longitudinal, helicity) or $\vec{S} \perp \vec{P}$ (transversal). In order to extract both spin structure functions from the tensor $W_{as}^{\mu\nu}$ one needs both polarizations. Let us consider first longitudinal polarization. Then we find in the Breit-frame

$$W_{as}^{12} = i \left[ g_1 - \left( \frac{M}{P_3} \right)^2 g_2 \right],$$
$$W_{as}^{21} = -W_{as}^{12}. \tag{40}$$

All other elements of $W_{as}^{\mu\nu}$ vanish. Now let us consider transversal polarization. Then we find in the Breit-frame

$$W_{as}^{02} = -i \frac{M}{P_3} (g_1 + g_2),$$
$$W_{as}^{20} = -W_{as}^{02}. \tag{41}$$

All other elements of $W_{as}^{\mu\nu}$ vanish again.

In the following let us consider the case where the proton as well as the partons are polarized longitudinally (helicity). We want to compute in the Breit-frame the 12-component of the hadronic tensor $W_{as}^{\mu\nu}$. We obtain

$$W_{as}^{12} = \frac{E}{4} \int d^2 k_\perp \sum_s \frac{l_{pol}^{12}(k_Q, k'_Q, s)}{Q\omega(\vec{k}_Q)} \left[ \Pi_s^b(P, \vec{k}_Q) + \Pi_s^d(P, \vec{k}_Q) \right], \tag{42}$$

where we have defined

$$l_{pol}^{12}(k_Q, k'_Q, s) = l_{\bar{u}u\bar{u}u}^{12}(k_Q, k'_q, s). \tag{43}$$

In the Breit-frame the anti-symmetric part of the leptonic tensor $l_{\bar{u}u\bar{u}u}^{\mu\nu}(k, k', s)$, Eq.(15), and of $l_{\bar{v}v\bar{v}v}^{\mu\nu}(k, k', s)$, Eq.(20), are given by

$$l_{\bar{u}u\bar{u}u}^{12}(k, k', s) = i2Qs_0,$$
$$l_{\bar{v}v\bar{v}v}^{12}(k, k', s) = -i2Qs_0. \tag{44}$$

For the parton spin $s$ holds $s \cdot k = 0$, $s^2 = -m^2$. In the helicity basis, one has $\vec{s} \parallel \vec{k}$. thus one defines the parton helicity $h$ by

$$h_s = \frac{\vec{s} \cdot \vec{k}}{\omega \mid \vec{k} \mid}. \tag{45}$$

In the Bjorken limes one obtains

$$\frac{l^{12}_{pol}}{Q\omega(\vec{k}_Q)} \to i2h_s. \tag{46}$$

and hence

$$W^{12}_{as} \to i\frac{P_3}{2} \int d^2k_\perp \sum_s h_s \left[ \Pi^b_s(P, \vec{k}_Q) + \Pi^d_s(P, \vec{k}_Q) \right], \tag{47}$$

Eq.(40) implies in the Bjorken limes

$$W^{12}_{as} \to ig_1. \tag{48}$$

From this and Eq.(47), after doing the same scale change as in the unpolarized case, we arrive at

$$g_1(x, Q) = \frac{1}{2} \sum_i e_i^2 \left[ h_+ f^{(i)}_+(x, 1, 2x\mu/Q) + h_- f^{(i)}_-(x, 1, 2x\mu/Q) \right.$$
$$\left. + h_+ \bar{f}^{(i)}_+(x, 1, 2x\mu/Q) + h_- \bar{f}^{(i)}_-(x, 1, 2x\mu/Q) \right]. \tag{49}$$

This is in agreement with the standard result of the parton model [28]

$$g_1(x) = \frac{1}{2} \sum_i e_i^2 \left[ h_+ q^{(i)}_+(x) + h_- q^{(i)}_-(x) + h_+ \bar{q}^{(i)}_+(x) + h_- \bar{q}^{(i)}_-(x) \right]. \tag{50}$$

In summary of this section, we have computed analytically in Breit-frame regularization the relation between hadronic tensor, structure functions and parton distribution functions. The main results are given in Eqs.(22,23) and Eqs.(42,43). The results are "renormalization group improved" compared to the naive parton model, taking into account parton mass $m$ and scale parameter $\mu$.

### D. Numerical results for distribution functions of the scalar model

In this section we want to show how distribution functions can be computed numerically with the Hamiltonian approach in the Breit-frame regularization. We apply the method to the scalar model in $3+1$ dimensions. This model has been extensively studied and represents, for finite cut-off, a non-trivial effective theory. We compute distribution functions for the $\phi^4$ and $\phi^3$ model. Let us firstly consider the $\phi^4_{3+1}$ model. The first excited state consists of only one parton (three-particle contributions have been found to be extremely small, i.e. within the error margin). This is because our particular choice of regularization (Breit-frame), which makes all parton momenta positive, implies for the Hamiltonian that those terms are dominant which conserve particle number. Consequently, the distribution function is peaked at $x_B = 1$, as shown in Fig.[3]. We did not find a noticeable dependence on the resolution $Q$, i.e., many-particle effects are absent. We have not put more effort to obtain a finer $x_B$ resolution, because this state does not display the interesting structure of a bound state. We have observed also that higher excited states display a dominant 2-, 3-, 4- particle content (with very small mixtures between different sectors). One should note, however, that the simplicity of the first excited state is due to the fact that the positivity of the longitudinal parton momenta prevents the creation of partons directly from the vacuum. Had we worked in the rest frame ($\vec{P} = 0$), the "valence parton" of the first excited state would be surrounded by a large cloud of partons with opposite momenta $\vec{p}$ and even the vacuum, lying in the $\vec{P} = 0$ sector, would be made up of such a cloud.

Now let us consider the $\phi^3$ model. We are aware that this theory is mathematically not well defined: It suffers from an unstable vacuum since it is unbounded from below. On the other hand, it is well defined in perturbation theory.

It is interesting to note that the $\phi^3$ model in $D = 6$ dimensions has the property of asymptotic freedom [18], opening the possibility to study perturbatively scaling violations. In $D = 4$ dimensions asymptotic freedom is known to exist only for non-Abelian gauge theories. This $\phi^3_{3+1}$ model serves here to illustrates how Altarelli-Parisi like behaviour and a sharp forward peak for small $x_B$ can be obtained in our method. In hadron physics one is ultimately interested in distribution functions of the proton which is a bound state. Its distribution functions reflect the strong variation of parton number of the proton and they look quite different from what we have seen above for the $\phi^4$ model. Thus in order to **illustrate** the capability of our method to treat such strong particle creation effects with a reasonable numerical effort, we have chosen the scalar $\phi^3$ model, which yields a bound state. We are aware that this theory is mathematically not well defined: It suffers from an unstable vacuum since it is unbounded from below. This unstable vacuum prevents a meaningful calculation of ground state masses which are needed to specify renormalization group trajectories and hence the exact relation between the resolution $Q$ and the bare coupling constant $g_0$. Therefore we can only compute the distribution function $\tilde{f}(x_B, g_0)$ where $g_0$ is the $\phi^3$ coupling constant. We have computed the distribution function in 1-, 2- and 3 space dimensions. For a given parton number cutoff these curves look very much alike. We present the result corresponding to a calculation in one space dimension (Fig.[4]) with $\Lambda/\Delta p = 11$ which implies that up to 11 partons can be present in the Fock space. When the coupling $g_0$ increases we see that the distribution function develops a peak at momentum fraction $x_B = 0$. This is because increasing the coupling means that more partons are produced which share the total momentum fraction. Decreasing the parton masses produces the same effect. Without a suitable regularization, the numerical effort to describe a system with a large particle number would be drastically higher, even for the $\phi^3$ theory.

## V. APPLICATION TO GAUGE THEORIES

Because the physically most important models are gauge theories we want to discuss the treatment of gauge theories in the Hamiltonian formulation with Breit-frame regularization. In the previous sections we have given arguments and numerical results showing the usefulness of a momentum lattice regularization in connection with the Breit-frame. The usefulness of a momentum lattice corresponding to the rest frame has been previously investigated and demonstrated by several workers: Kuti and co-workers [19] have investigated the one-component scalar $\phi^4$ model and the $O(4)$ symmetric scalar model and estimated a bound on the Higgs mass. Kröger and co-workers [20] have solved the Langevin equation on a momentum lattice for the scalar $\phi^4_{3+1}$ model and extracted critical behavior. Glueball scattering in compact $QED_{2+1}$ ($QCD$-like model) has been computed on a momentum lattice in Ref. [16]. Properties of nuclear matter have been computed by Brockmann and Frank [21]. Kogut and co-workers [22] have studied the phase diagram of quenched $QED$ on a momentum lattice. Espriu [23] has studied the renormalization group flow by use a momentum lattice. Finally, Koutsoumbras [24] has computed the gluon propagator of finite temperature $QCD$ from a momentum lattice. Thus a momentum lattice regularization has proven useful when studying numerically physics near a critical point.

When one treats gauge theories on a momentum lattice the following problem occurs: If one takes the gauge fields $A_\mu(k_i)$ as variables (so-called non-compact formulation), where $k_i$ denotes a momentum lattice, then the gauge action is not manifestly gauge invariant. As consequence one has observed non-local counter terms when computing from lattice perturbation theory the axial anomaly and the one-loop vacuum polarization. This has been seen by Karsten and Smit [37] by computing the triangle diagram using the SLAC derivative in the action and by Kröger and co-workers [38] using an action defined on a momentum lattice with a momentum cut-off $\Lambda$. As Wilson has pointed out, it is desirable to conserve gauge symmetry manifestly in a regularized gauge theory. E.g., there is numerical evidence [39] that a lattice action being not manifestly gauge invariant yields no area law for the Wilson loop in pure $SU(2)$ gauge theory.

The space-time lattice Hamiltonian, corresponding to the Wilson action and being manifestly gauge invariant, has been constructed by Kogut and Susskind [40]. Here we are confronted with the following problem: How to introduce a momentum lattice as regulator while conserving manifestly gauge invariance? We suggest to do this as follows: We consider as variables closed Wilson-loops (for pure gauge theory without color charges). The Hilbert space is built from those loops. Gauge invariance corresponds to fulfilling Gauss' law

$$G \mid \phi > = 0. \tag{51}$$

For a fixed lattice site $i$, one has $G_i = \sum_{\{i,j\} \ni i} l^a_{ij}$, i.e., the sum over generators of gauge transformations (where the temporal gauge is fixed). States corresponding to closed loops obey this law, while states corresponding open strings do not obey it. Physical states are color singlet states, thus open string states are unphysical. Nevertheless we will make use of them as an intermediate step when constructing the Hilbert space of states obeying the Breit-condition.

In order to introduce a regularization, we start from a conventional space-time lattice (regular, hypercube) with lattice spacing $a$. Then closed loops as well as open strings are defined as curves connecting adjacent lattice sites (straight line between neighboring lattice sites). E.g., a loop state is given by

$$| \phi(x_1) >= | U_\mu(x_1) U_\nu(x_1 + a\hat{\mu}) \cdots U_\omega(x_N) >, \qquad (52)$$

where $x_N + a\hat{\omega} = x_1$. In order to introduce a momentum lattice we make a discrete Fourier transformation,

$$| \tilde{\phi}(k_i) >= a \sum_{x_i} \exp[-ix_i k_i] | \phi(x_i) >, \qquad (53)$$

where each component of $k_i$ runs over the Brillouin zone $-\pi/a$ to $+\pi/a$. One can define the lattice momentum operator $P_\mu$ via the lattice translation $T_\mu(a)$, which translates each configuration on the lattice by an increment $a$ in the direction $\mu$. It is given by

$$T_\mu(a) = \exp[-iaP_\mu]. \qquad (54)$$

The eigenvalues of $P$ are $k_i$, which are the possible momenta of the loop state.

In order to construct states with well defined momentum, obeying the Breit-condition, plus satisfying gauge invariance, we suggest to proceed as follows: We construct a Hilbert space built from link states. By discrete Fourier transformation we associate a discrete lattice momentum to each link, say $\tilde{U}_\mu(k_i)$. Then we construct multiple link states

$$| \psi(k_1, \cdots, k_N) >= | \tilde{U}_\mu(k_1) \tilde{U}_\nu(k_2) \cdots \tilde{U}_\omega(k_N) > . \qquad (55)$$

This state corresponds to momentum $k_{tot} = k_1 + \cdots k_N$. Then we impose the Breit-condition (see Eq.(4))

$$(\vec{k}_i - \vec{P}/2)^2 \leq (\vec{P}/2)^2, \qquad (56)$$

where the parton momenta are given by the lattice momenta of the links. Thus, like in the scalar model, positivity of parton momenta in all three components and a given total momentum limits the total number of links to $\frac{|\vec{P}|}{\Delta p}$ and thus give drastic bounds on the dimension of the effective Hilbert space. Eventually, we implement gauge symmetry by requiring Gauss' law, Eq.(51), to be respected. Thus the Breit-condition and Gauss' law define our basis of Hilbert states.

## VI. $S$-MATRIX

The Hamiltonian in the Breit-frame regularization has been shown above to be a suitable tool in the scalar model for computation of the mass spectrum and physics at the critical line as well as distribution functions. In this part we want to suggest that it is also a valuable tool for scattering phenomena and in particular for the non-perturbative computation of the $S$-matrix. When considering non-perturbative computation of scattering observables, standard Euclidean lattice field theory is faced with the following problem: Scattering matrix elements are directly related to Minkowski n-point functions. On the lattice one can compute Euclidean n-point functions. In principle there is an analytic continuation between those two types of n-point functions. However, when the Euclidean n-point function is only known at some lattice points within the uncertainty of statistical errors, it is very difficult (almost impossible) to get reliable numerical results from an analytic continuation. A way out of this dilemma has been proposed by Lüscher [29]. The idea is that continuum scattering phases can be extracted from the finite-size behavior of a mass spectrum on a finite lattice. This requires mass calculations via standard Euclidean lattice techniques, but requires quite precise data in order to resolve finite size effects.

An alternative way to compute non-perturbatively an $S$-matrix has been suggested by Kröger and co-workers [17]. The idea is the following. The $S$-matrix, as has been introduced by Heisenberg [30] and Møller [31], is defined as

$$S = <\psi^{(-)} | \psi^{(+)}>, \tag{57}$$

which is the probability amplitude to find an outgoing scattering state in an incoming scattering state. The scattering states are characterized by two conditions: (i) they are eigenstates of the Hamiltonian, and (ii) for $t \to \pm\infty$ they approach an asymptotic state. The asymptotic state describes two non-interacting particles (in the case of two-particle scattering). The so-called Møller operator maps the asymptotic states $|\phi^{as}>$ onto the scattering states $|\phi^{(\pm)}>$.

$$|\phi^{(\pm)}> = \Omega^{(\pm)} |\phi^{as}> = s - \lim_{t \to \mp\infty} \exp[iHt]\exp[-iH^0 t]\phi^{as} > . \tag{58}$$

Those equations define scattering states and the $S$-matrix. They can be taken over to quantum field theory with some care.

### A. Asymptotic one- and two-particle states

One problem is the construction of an asymptotic one-particle state, asymptotic two-particle state, etc. In constructive quantum field theory this is answered by Haag-Ruelle theory [32]. It tells how to construct asymptotic one-particle states by application of suitable local field operators on the physical vacuum.

$$|1>_{phys} = a^\dagger(f) |0>_{phys}, \tag{59}$$

and a two-particle state is given by

$$|2>_{phys} = a^\dagger(f_1) a^\dagger(f_2) |0>_{phys} . \tag{60}$$

Here $a^\dagger$ is the creation operator of a one-particle state with wave function $f$ created from the physical vaccum. The existence of such an operator has been proven by Haag and Ruelle [32]. An explicit form of this operator for the case of glueball states in pure gauge theory has been given by Lüscher [33]. But Haag-Ruelle theory says nothing about how to find the physical vacuum. In the Hamiltonian approach in connection with the Breit-frame regularization, as advocated here, we avoid constructing the physical vacuum. Thus we proceed a route alternative to Haag-Ruelle's theory. We construct a one-particle state with momentum $\vec{p}$ directly by calculating an eigenvector of the regularized Hamiltonian $H$,

$$H |\vec{p}> = E(\vec{p}) |\vec{p}> . \tag{61}$$

The property of being a one-particle state is verified by computing its mass (see sect.3). If, e.g., its mass is the lowest mass of the mass spectrum, the state $|\vec{p}>$ is a one-particle state. Let $|f>$ denote such a one-particle state with momentum distribution given by a wavefunction $f$. In the language of Haag-Ruelle theory, the explicit construction of the state $|f>$ means that we have found a creation operator $A^\dagger(f)$ with

$$|1> = A^\dagger(f) |0>_{free} . \tag{62}$$

I.e., it creates a one-particle state from the vacuum of the regularized free Hamiltonian. There is a theorem by Haag [34] which says that in the continuum limit of relativistic quantum field theory, the physical Hilbert states of the interacting field theory (Hamiltonian) have nothing to do with those of the free field theory (free Hamiltonian). In particular there is no unitary transformation between the physical vacuum to the free vacuum. However, this theorem does not apply when we consider the *regularized* field theory (Hamiltonian). Then there is a unitary transformation $U$, mapping the (regularized) free vacuum onto the (regularized) physical vacuum,

$$|0>_{phys}^{reg} = U |0>_{free}^{reg} . \tag{63}$$

This relates the Haag-Ruelle creation operator $a^\dagger(f)$ (of the regularized field theory) to the creation operator $A^\dagger(f)$ by

$$a^\dagger(f) = U A^\dagger(f) U^{-1} . \tag{64}$$

Finally, using $A^\dagger(f)$ from Eq.(42) we can construct in equivalence to Eq.(40) asymptotic non-interacting two-particle states by

$$|2> = A^\dagger(f_1) A^\dagger(f_2) |0>_{free} . \tag{65}$$

## B. Møller wave operators and $S$-matrix

Let us denote by $\mid \phi^{as}(\vec{p}_1, \vec{p}_2) >$ the asymptotic two-particle state, corresponding to two non-interacting particles with momentum $\vec{p}_1$ and $\vec{p}_2$ respectively. Then the Møller wave operator is given by

$$\mid \phi^{(\pm)}_{scatt}(T) >= \Omega^{(\pm)}(T) \mid \phi^{as}(\vec{p}_1, \vec{p}_2) >= \exp[\mp iHT] \exp[\pm i(E(\vec{p}_1) + E(\vec{p}_2))T] \mid \phi^{as}(\vec{p}_1, \vec{p}_2) > . \qquad (66)$$

Here $E(\vec{p})$ denotes the energy-momentum dispersion relation of the one-particle state of mass $m$. $H$ denotes the regularized Hamiltonian. The time parameter $t$, which goes to infinity in the continuum limit, has to be chosen to take a positive finite value $T$ in the regularized theory (see below). In a similar way, one can construct the $S$-matrix

$$S_{fi,in}(T) =< \phi^{as}_{fi} \mid \exp[i(E(\vec{p'}_1) + E(\vec{p'}_2))T] \exp[-i2HT] \exp[i(E(\vec{p}_1) + E(\vec{p}_2))T] \mid \phi^{as}_{in} > . \qquad (67)$$

From the numerical point of view, the computation of the $S$-matrix element proceeds most simply by diagonalizing the regularized Hamiltonian

$$H \mid \eta_\nu >= \epsilon_\nu \mid \eta_\nu >, \quad \nu = 1, 2, \cdots$$
$$\exp[iHT] = \sum_\nu \mid \eta_\nu > \exp[i\epsilon_\nu T] < \eta_\nu \mid . \qquad (68)$$

How should one choose the scattering time parameter $T$? When applying this time-dependent Hamiltonian method to non-relativistic quantum mechanics as well as to field theoretic models [17], the following general observations have emerged from numerical calculations: The matrix element $S_{fi,in}(t)$ considered as a function of $t$ has the following behavior. At $t = 0$ it takes the value $< \phi^{as}_{fi} \mid \phi^{as}_{in} >$ (in the case of elastic scattering). When increasing $t$ it deviates from the starting value and eventually reaches a plateau region. When further increasing $t$, it leaves the plateau region and after a while shows an (irregular) oscillatory behavior. The plateau region is the region of physical interest. Its existence can be shown analytically for non-relativistic potential scattering (see Ref. [17] and references therein). The location and size of this plateau region depends on the model and dimension. In particular, it depends on the dimension of the regularized Hamiltonian. When increasing this dimension, i.e., when exploring a larger Hilbert space, the size of the plateau region becomes larger. In the continuum limit, when the $S$-matrix converges, the size of the plateau should become infinitely large. The time parameter $T$ should be chosen from this plateau region, either by determining where the matrix element $S_{fi,in}(t)$ has the least variation in $t$, or by the following criterion of conservation of energy: In the continuum limit, energy conservation in a scattering reaction means that

$$< \psi^{(\pm)} \mid H \mid \psi^{(\pm)} >= E_{as}, \qquad (69)$$

where $E_{as}$ denotes the energy of the asymptotic non-interacting two-particle state. Thus we define the function

$$\Delta E(t) = \left| < \Omega^{(\pm)}(t) \phi^{as}(\vec{p}_1, \vec{p}_2) \mid H \mid \Omega^{(\pm)}(t) \phi^{as}(\vec{p}_1, \vec{p}_2) > -E_{as} \right| / E_{as}, \qquad (70)$$

where $\Omega^{(\pm)}$ is given by Eq.(58). This function is a measure of violation of energy conservation in a scattering reaction computed with the regularized Hamiltonian at some finite time $t$. In the continuum limit this should be zero. Thus we can choose the time paramer $T$ such that $\Delta E(t)$ has a minimum. Numerical experience has shown that the value of $T$ once determined as position of minimal variation of the $S$-matrix element and secondly determined as position of the minimum of $\Delta E$ are quite close together, which is an indication of consistence.

In order to get the physical $S$-matrix one has to carry out renormalization and take into account the vacuum structure. Renormalization means that firstly one has to determine the counter terms in the Hamiltonian. E.g., for the scalar $\phi^4$ model, one has to renormalize the wavefunction, the mass and the coupling constant. Then one computes physical observables like, e.g., masses or scattering cross sections and tunes the bare parameters of the model, such that the physical observables remain fixed. Finally, the vacuum structure needs some careful treatment. The computation of the $S$-matrix, as described above yields the full $S$-matrix, which includes the connected part (which is the piece observed in scattering experiments) but also all disconnected parts. The factorization of n-point Greens functions into connected pieces, is knows as vacuum structure [35]. This allows to extract the connected part of the $S$-matrix.

The time-dependent Hamiltonian method as described above, but instead of using the Breit-frame regularization by use of the rest-frame regularization, has been applied to glueball scattering in compact $U(1)$ gauge theory (compact $QED$) in $2 + 1$ dimensions [16]. At the end of this section we want to adress the question: What advantage does it bring to use the Breit-frame regularization for scattering calculations in the time-dependent Hamiltonian formulation? Firstly, as mentioned above, the Breit-frame regularization avoids the calculation of the vacuum state when constructing asymptotic non-interacting two-particle states. Secondly, this regularization reduces the number of effective degrees of freedom by the same mechanism as was shown to be useful for the calculation of the mass spectrum. However, one must pay attention to the following limitation: Because we take into account parton momenta inside the sphere given by the Breit condition Eq.(4), the momenta of the asymptotic particles, i.e., $\vec{p}'_1, \vec{p}'_2, \vec{p}_1, \vec{p}_2$ should lie well inside the Breit sphere. This limits the scattering reactions which can be treated. E.g., head-on collisions are not included. However, this constraint is not very stringent, because a suitable Lorentz-boost can be applied to map the momenta into the Breit sphere.

## VII. FINITE DENSITY THERMODYNAMICS

The computation of thermodynamic observables at finite temperature and finite density is an important problem for the physics of neutron stars, high energy heavy ion collisions, and for the question of phase transition from the hadronic phase to quark-gluon plasma in $QCD$. However, when treating finite temperature $QCD$ in the standard Lagrangian lattice approach, there is an well known problem when a non-zero chemical potential is included to describe finite density effects. Then the fermionic determinant becomes complex yielding a complex lattice action. This has led to great difficulties when solving the model numerically via Monte Carlo methods [36]. In order to study the infrared dynamics of Yang-Mills and Yang-Mills-Higgs theories at finite temperature, which can not be addressed by Euclidean methods, Moore [41] has suggested an improved Hamiltonian for Minkowski Yang-Mills theory.

In this section we want to discuss how finite temperature and finite density thermodynamics can be treated in a Hamiltonian formulation with the Breit frame regularization. The point is that the Hamiltonian formulation allows to treat also non-Hermitian Hamiltonians (complex actions). Consider the following partition function

$$Z = Tr \exp[-\frac{1}{k_B T}(H_0 + \mu N)], \tag{71}$$

where $H_0$ stands for a Hermitian Hamiltonian, $\mu$ denotes the chemical potential and $N$ stands for a particle number operator. Let us suppose now, for the sake of the argument that the term $\mu N$ would be non-Hermitian. What is then the advantage to use a Hamiltonian formulation? In a Hamiltonian formulation this partition function can be computed non-perturbatively via diagonalization of $H_0 + \mu N$ in the same way as $exp[iHt]$ has been computed when calculating the $S$-matrix (sect.6).

What is the advantage to use the Breit-frame regularization? Let us consider the following scenario: One wishes to study hot nuclear matter, respectively quark-gluon plasma in a state with total momentum $\vec{P} \neq 0$. In order to investigate this experimentally, one can perform a high energy heavy ion collision experiments with large momentum $\vec{P}$. However, it is stil an open question if thermodynamical equilibrium can be reached during the short time of collision and before decaying of fragments, which would justify the use of the Boltzmann -Gibbs partition function. Putting aside for the moment the question of experimental realization, it is nevertheless physically interesting to ask the following question: What are the properties of matter at finite temperature and finite density at thermodynamical equilibrium in a sector of momentum $\vec{P} \neq 0$. Thus we consider the partition function at momentum $\vec{P}$,

$$Z(\vec{P}) = Tr \left[ \exp[-\frac{1}{k_B T}(H_0 + \mu N)] \right]\bigg|_{\vec{P}}. \tag{72}$$

The evaluation of this function now can be done in the Breit-frame regularization, which by the same reason as in the computation of structure functions would reduce the effective number of degrees of freedom, i.e., the dimension of the effective Hilbert space.

## VIII. SUMMARY

In conclusion, we have suggested a Hamiltonian method and a momentum regularization corresponding to the Breit-frame. We have shown that this method allows to extract continuum physics by presenting numerical results for the $\phi^4_{3+1}$ theory in the symmetric phase close to the critical line. We find close agreement with the solution of the renormalization group equations by Lüscher and Weisz. We have seen scaling behaviour of several low-lying masses near the critical point. Using the Breit-frame, we have computed analytically for $DIS$ in $QCD$ the relation between the hadronic tensor, the structure functions and the quark distribution functions. In the Bjorken limes we find the conventional relations between $F_1$, $F_2$, $g_1$ and the quark distribution functions. We have presented numerical results for parton distribution functions for the $\phi^4$ and $\phi^3$ model. The example of the $\phi^3$ model demonstrates how a peak at small $x_B$ can be produced, with a behavior similar to Altarelli-Parisi behavior in $QCD$. We have proposed how the Breit-frame regularization can be applied to gauge theories, while keeping gauge symmetry manifestly conserved. We have suggested that this regularization might be useful also for the computation of scattering reactions ($S$-matrix), as well as finite temperature and finite density thermodynamics. We are optimistic that the method can be applied to compute numerically structure functions in $QCD$. Work is in progress.

## ACKNOWLEDGMENTS

H. Kröger gratefully acknowledges support by NSERC Canada. N.Scheu wants to express his appreciation for having been granted the AUFE fellowship from the DAAD (Deutscher Akademischer Austauschdienst) which has made this Ph.D. project possible. The authors are grateful for discussions with D. Schütte.

**Fig.1** The ground state mass $m_R$ in lattice units ($a \equiv 1$) versus $\kappa$ for $\lambda = 0.00345739$ ($\bar{\lambda} = 0.01$ in Ref. [2]), $\lambda$ and $\kappa$ are given by Eq.(6). The points correspond to results of Ref. [2]. Our results correspond to $\Lambda/\Delta p = 3$ (dashed line) and $\Lambda/\Delta p = 4$ (solid line).

**Fig.2** The lowest lying mass spectrum versus $\kappa$. The ground state mass is set to one, $\lambda$ as in Fig.[1]. $\lambda$ and $\kappa$ are given by Eq.(6). $\Lambda/\Delta p = 4$.

**Fig.3** The distribution function $f(x_B)$ of $\phi^4_{3+1}$ versus the momentum fraction $x_B$. $\lambda = 0.00345739$ (as in Fig.[1]), $\Lambda/\Delta p = 4$.

**Fig.4** The distribution function $\tilde{f}(x_B, g_0)$ of $\phi^3_{1+1}$ versus the momentum fraction $x_B$ and the coupling constant $g_0$. The bare mass $m_0$ has been set to $m_0 = 3\Delta k$. $\Lambda/\Delta p = 11$.

## IX. TABLE CAPTION

| $\lambda$ | 0.0005 | 0.001 | 0.005 | 0.01 | 0.05 | 0.1 |
|---|---|---|---|---|---|---|
| $\kappa_{crit}^{LW}$ | 0.125101 | 0.125202 | 0.125991 | 0.126968 | 0.132368 | 0.13601 |
| $\alpha$ | 0.99997 | 0.99993 | 0.99972 | 0.9993 | 1.0073 | 1.0275 |

The critical points $\kappa_{crit}$ versus $\lambda$. $\lambda$ and $\kappa$ are given by Eq.(6). $\kappa_{crit}^{LW}$ is taken from Ref. [2]. $\alpha := \kappa_{crit}^{KS}/\kappa_{crit}^{LW}$ denotes the ratio between the results of this work and Ref. [2]. In this work, $\kappa_{crit}$ has been determined under the condition that the renormalized mass $m_R$ becomes imaginary. $\Lambda/\Delta p = 4$.

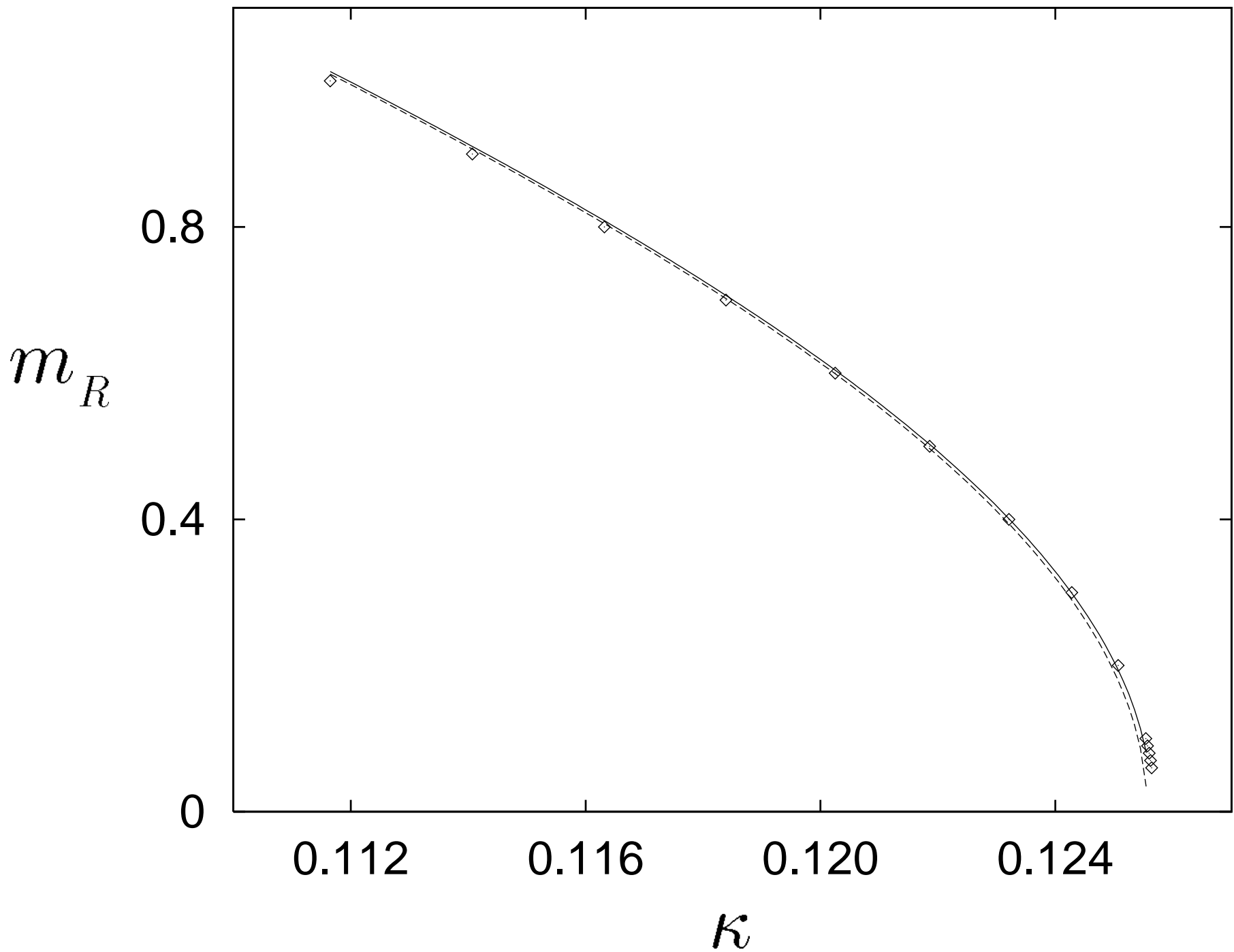

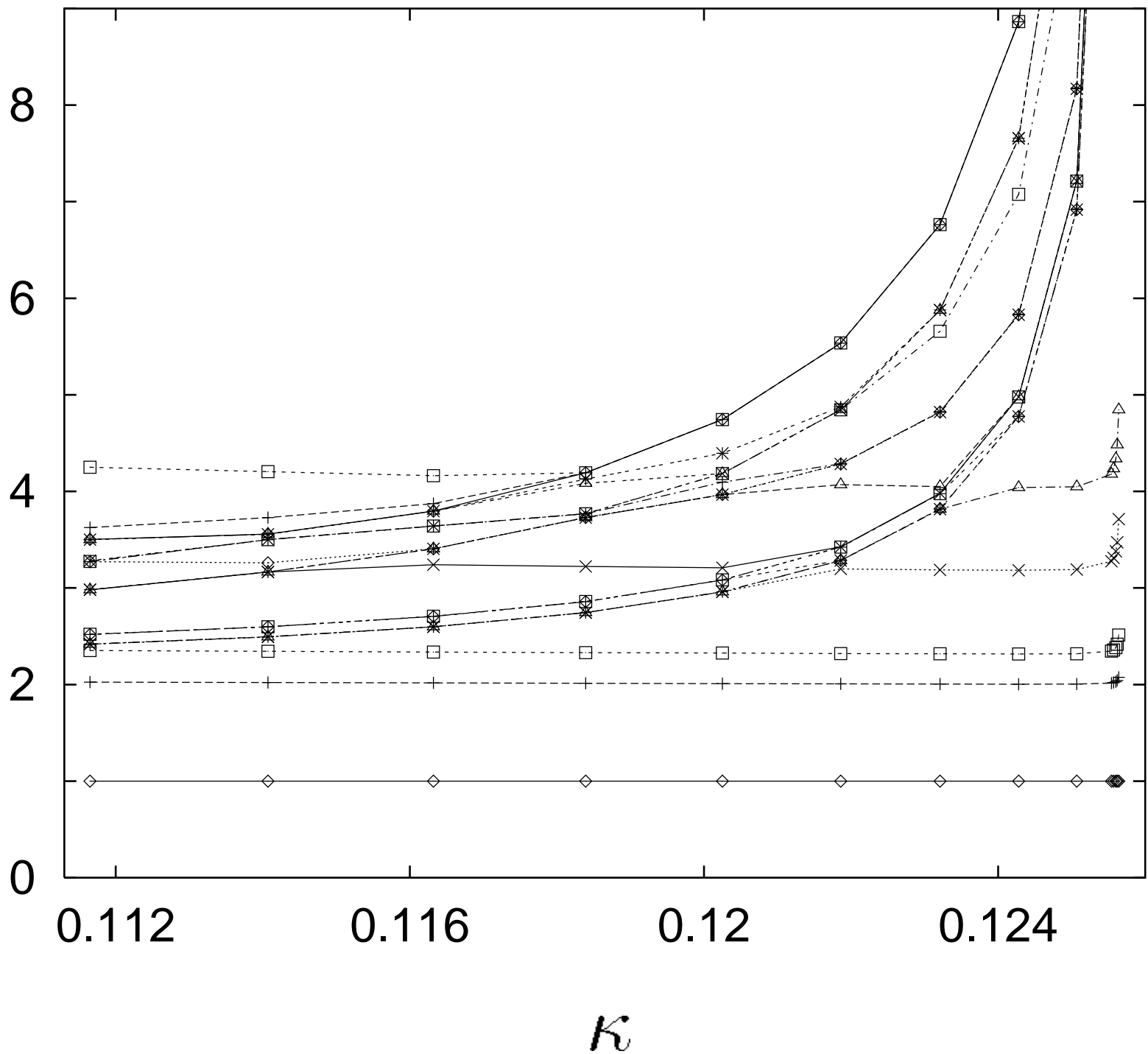

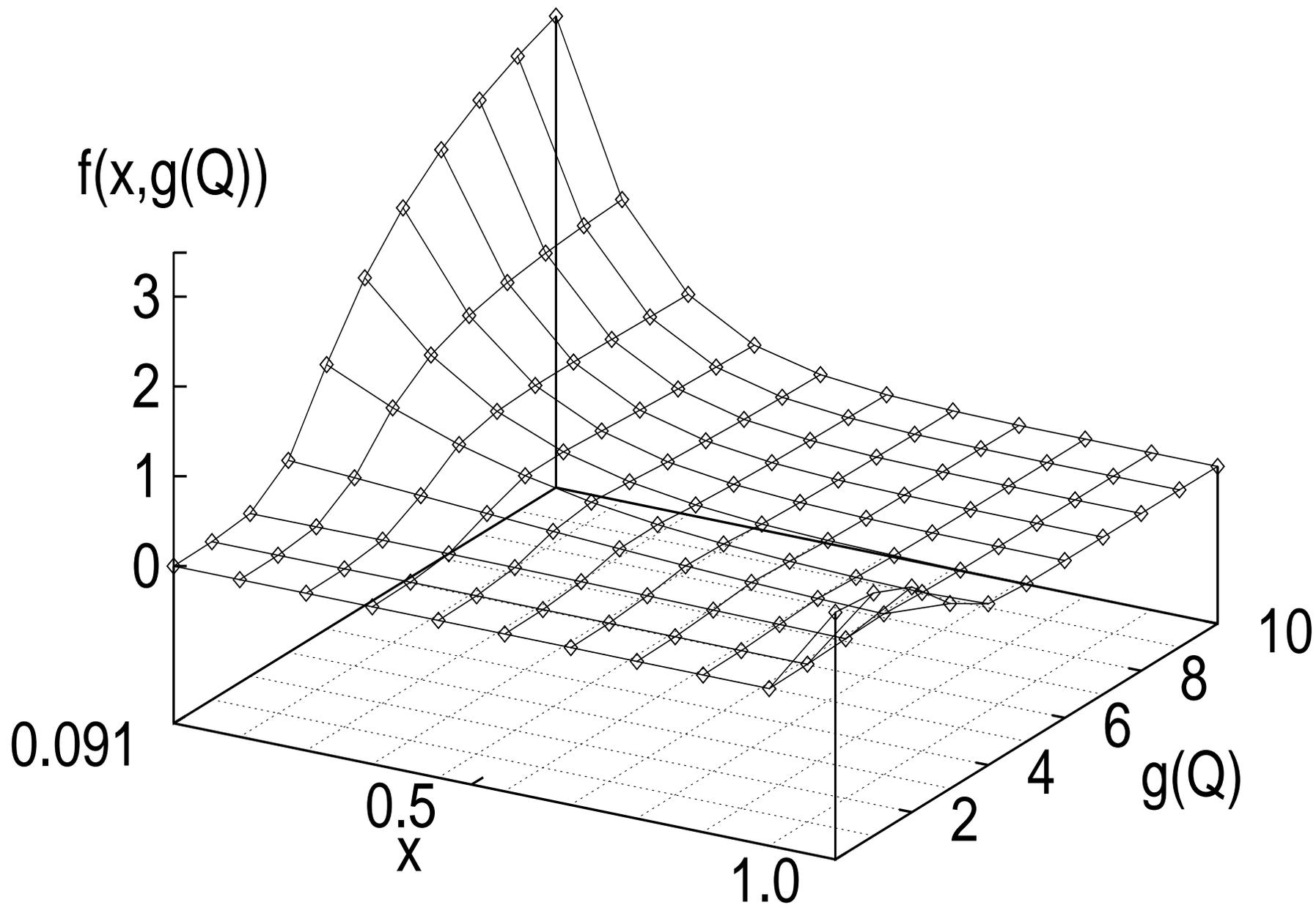

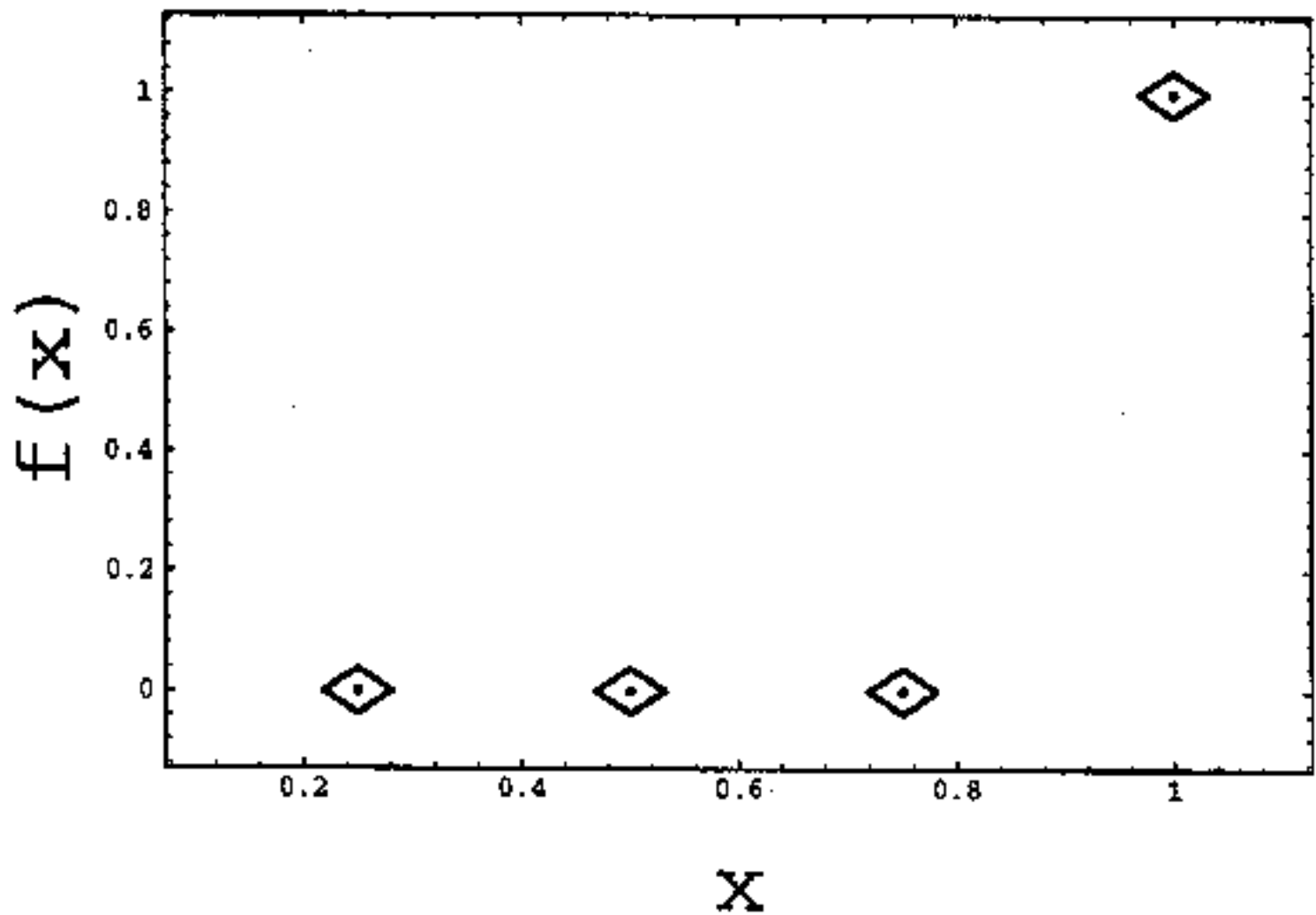